\documentclass{svproc}
\usepackage{graphicx}%
\usepackage{multirow}%
\usepackage{amsmath,amssymb,amsfonts}%
\usepackage{mathrsfs}%
\usepackage[title]{appendix}%
\usepackage{xcolor}%
\usepackage{textcomp}%
\usepackage{manyfoot}%
\usepackage{booktabs}%
\usepackage{algorithm}%
\usepackage{algorithmicx}%
\usepackage{algpseudocode}%
\usepackage{listings}%
\usepackage{url}

\usepackage{subcaption}

\usepackage{CJKutf8}

\usepackage[breaklinks]{hyperref}

\begin{document}
\mainmatter              
\title{Network Contagion in Financial Labor Markets: Predicting Turnover in Hong Kong}
\titlerunning{Turnover Prediction in Hong Kong’s Financial Network}  
%
\author{Abdulla AlKetbi\inst{1,2,3} \and Patrick Yam\inst{3} \and Gautier Marti\inst{3} \and Raed Jaradat\inst{1}}
\authorrunning{Abdulla AlKetbi et al.} 

\institute{Khalifa University, Shakhbout Bin Sultan St - Hadbat Al Za'faranah - Abu Dhabi,\\
\email{100061314@ku.ac.ae},\\ WWW home page:
\texttt{https://www.ku.ac.ae}
\and
ADIA Lab, Level 26, Al Khatem Tower, ADGM Square, Al Maryah Island,\\
Abu Dhabi, United Arab Emirates
\and
Abu Dhabi Investment Authority (ADIA)
}

\maketitle              
\begin{abstract}
Employee turnover is a critical challenge in financial markets, yet little is known about the role of professional networks in shaping career moves. Using the Hong Kong Securities and Futures Commission (SFC) public register (2007–2024), we construct temporal networks of 121,883 professionals and 4,979 firms to analyze and predict employee departures.
We introduce a graph-based feature propagation framework that captures peer influence and organizational stability. Our analysis shows a contagion effect: professionals are 23\% more likely to leave when over 30\% of their peers depart within six months. Embedding these network signals into machine learning models improves turnover prediction by 30\% over baselines.
These results highlight the predictive power of temporal network effects in workforce dynamics, and demonstrate how network-based analytics can inform regulatory monitoring, talent management, and systemic risk assessment.
\end{abstract}

\section{Introduction}\label{sec1}

The financial services sector is central to Hong Kong’s economy, employing over 270,000 professionals and contributing 23\% of GDP \cite{hkma_ifc_website}. Yet beyond macroeconomic indicators, the micro-dynamics of professional mobility remain poorly understood. Employee turnover in particular represents both a systemic risk and a major operational cost, with annual rates exceeding 24\%.\footnote{\href{https://www.fsdc.org.hk/media/52ulbcmf/20230824_-en-nurturing-the-young-talents-of-today-for-hk-s-fs-industry-of-tomorrow_fv.pdf}{FSDC Report (2023)}}

Recent work has begun to map Hong Kong’s financial ecosystem through the Securities and Futures Commission (SFC) register, revealing its large-scale network of professionals and firms \cite{alketbi2024mapping}. In this paper, we move from mapping to forecasting: can we predict when a professional will leave their firm, and does the surrounding network structure improve predictive power?

To address this, we construct temporal networks from 2007–2024, capturing career trajectories of 121,883 professionals across 4,979 firms. We develop a graph-based feature propagation framework to quantify peer influence and organizational stability, and integrate these signals into predictive models. Our approach tackles four challenges related to workforce analytics: 

\begin{enumerate}
    \item \textbf{Network contagion}: capturing peer turnover cascades through temporal propagation;
    \item \textbf{Temporal dynamics}: evaluating models with strict walk-forward validation;
    \item \textbf{Class imbalance}: rare-event prediction with only 2.3\% monthly turnover;
    \item \textbf{Feature engineering}: combining individual, organizational, and propagated network features.
\end{enumerate}

We show that network-derived signals substantially improve predictive accuracy over non-network baselines, and that firm-level stability and peer influence dominate demographic attributes. These findings highlight the role of contagion in financial labor markets and open the door to network-based workforce analytics with applications in recruiting, talent management, systemic risk assessment, and even the design of immigration policies for skilled workers.


\section{Related Work}\label{sec:related}

Our study connects four strands of research: (i) network and contagion mechanisms of employee turnover, (ii) temporal and diffusion methods from complex systems, (iii) digital-trace approaches to labor mobility, and (iv) machine-learning frameworks for attrition prediction.

\paragraph{Turnover and organizational networks.}
Seminal work established that relational embeddedness predicts exits beyond individual traits: employees tied to leavers or located on network peripheries are more likely to depart \cite{feeley2008predicting}. Vardaman et al.\ further showed that network structure interacts with job embeddedness, especially \emph{Simmelian ties} (strong ties embedded in triads that exert reinforcing social pressure), to determine whether turnover intentions translate into actual exits \cite{vardaman2015translating}. More recently, organizational network analysis has been operationalized for HR retention, with churn prediction models leveraging centrality, k-core structure, and cohesion to flag at-risk employees \cite{younis2023employee}. Our study is aligned with this tradition but differs in scale and source: we exploit a \emph{regulatory registry} covering 121{,}883 individuals across 4{,}979 firms, and we quantify contagion through \emph{temporal feature propagation} rather than static ego-network metrics.

\paragraph{Temporal and diffusion processes in complex networks.}
From the complex networks community, temporal networks provide the toolkit for constructing time-resolved graphs, measuring evolving connectivity, and avoiding leakage across time \cite{holme2012temporal,masuda2016guide}. Contagion and diffusion models, originally developed for epidemics and behavioral adoption, clarify how peer influence can exhibit thresholds and reinforcement (``complex contagion'') \cite{pastor2001epidemic,centola2010spread}. We operationalize these ideas for labor mobility by creating monthly snapshots and propagating turnover signals over employee and firm graphs, thereby testing whether departures diffuse through professional ties.

\paragraph{Digital traces and labor mobility.}
Large-scale online platforms have also been mined for workforce mobility. State et al.\ analyzed LinkedIn career histories to study the migration of professionals to the United States, revealing macro-level flows of talent across countries \cite{state2014migration}. Perrotta et al.\ extended this line of work by using LinkedIn Recruiter data to measure Europeans’ openness to international migration, combining descriptive flow analysis with gravity-type models \cite{perrotta2022openness}. While valuable for descriptive mapping of aspirations and flows, LinkedIn data are self-reported, can be retrospectively edited, and typically lag real events. In practice, it is \emph{users} who strategically back-populate their profiles—sometimes months or even years later—whether to mask involuntary exits, to wait until a trial period is completed, or to gauge whether a new role is worth keeping before making it public. These properties make LinkedIn less suitable for \emph{live prediction} of churn. By contrast, our regulator-sourced registry provides point-in-time accuracy with legal force, enabling forward-looking attrition forecasting rather than retrospective mapping.

\paragraph{ML for attrition and network-aware churn.}
In parallel, the machine learning literature has progressed from classic classifiers toward network-aware models. Recent evidence in multilayer settings shows that network embeddedness substantially improves prediction and supports explanation (e.g., SHAP-style attribution) compared to attribute-only baselines \cite{gadar2024explainable}. Related studies using internal communication or collaboration networks similarly report gains when network features are included. Our contribution is to bring these network-aware ML insights to a regulated financial labor market, using a comprehensive regulator-sourced registry that covers the entire Hong Kong financial center, with strict walk-forward validation. Methodologically, we contribute a lightweight \emph{graph-based feature propagation} framework that yields complementary signals to individual and firm covariates, and we document when and where such signals are most predictive.

\paragraph{Positioning and novelty.}
Relative to prior work, we (i) shift from single-firm ONA or digital-trace platforms to market-wide, regulator-sourced temporal networks; (ii) formalize peer effects with an explicit propagation operator over monthly snapshots rather than relying only on static centralities; (iii) evaluate predictability \emph{out-of-sample} across 169 months with class-imbalance–robust metrics; and (iv) show that firm stability and propagated peer effects dominate demographics (LLM-inferred) in feature importance. Together, these choices bridge complex network contagion and practical turnover forecasting in a high-stakes, regulated domain.

\section{Dataset and Preprocessing}\label{sec:dataset}

\subsection{Origin of the Dataset}
This study relies on the \textit{Public Register of Licensed Persons and Registered Institutions} maintained by the Hong Kong Securities and Futures Commission (SFC).\footnote{\url{https://www.sfc.hk/en/Regulatory-functions/Intermediaries/Licensing/Register-of-licensed-persons-and-registered-institutions}} The register has systematically recorded licensed individuals and their employing firms since the implementation of the Securities and Futures Ordinance in 2003. It provides detailed information on employment periods, firm affiliations, and regulated activities, covering the core of Hong Kong’s financial workforce. 

The dataset used in this work spans April 2003 to March 2024 and contains 519{,}860 employment records across 121{,}883 unique professionals and 4{,}979 firms. A first descriptive analysis of this register, focusing on the structure of the financial ecosystem, was presented in our previous work \cite{alketbi2024mapping}. Here we extend this line of research by constructing temporal networks to study and forecast employee turnover.

\subsection{Temporal Structure and Target Variable}
For predictive modeling, we create 206 monthly snapshots from 2007--2024, aligning all employment spells to a common temporal grid. At each month $t$, we record whether a licensee’s employment ends in the subsequent month. This defines the binary target variable \textit{turnover}, with a positive rate of only 2.3\% per month (27.6\% annual rate). This base rate of 2.3\% underscores the difficulty of the prediction task and motivates specific strategies for class imbalance.

\subsection{Feature Engineering Overview}
We engineer four broad categories of features from the raw SFC register:

\begin{itemize}
    \item \textbf{Individual-level features}: career history (number of firms, mobility rate), tenure at the current employer, and licensing role (Representative or Responsible Officer).
    \item \textbf{Firm-level features}: organizational size (number of active employees), stability (average and quantile employee tenure), and longevity (firm age).
    \item \textbf{Network-propagated features}: signals derived from employee--employee and firm--firm networks, capturing peer turnover rates, organizational contagion, and industry-wide stability.
    \item \textbf{Demographic features}: inferred from names using large language models (LLMs), following the approach of \cite{alnuaimi2024enriching}. This enrichment provided likely gender and country-of-origin attributes, allowing us to test whether demographic factors contribute to turnover prediction.
\end{itemize}

Altogether, this results in 36 features per record, combining static descriptors with dynamic, network-based signals.

\subsection{Descriptive Statistics}
\begin{figure}[h]
\centering
\includegraphics[width=0.9\linewidth]{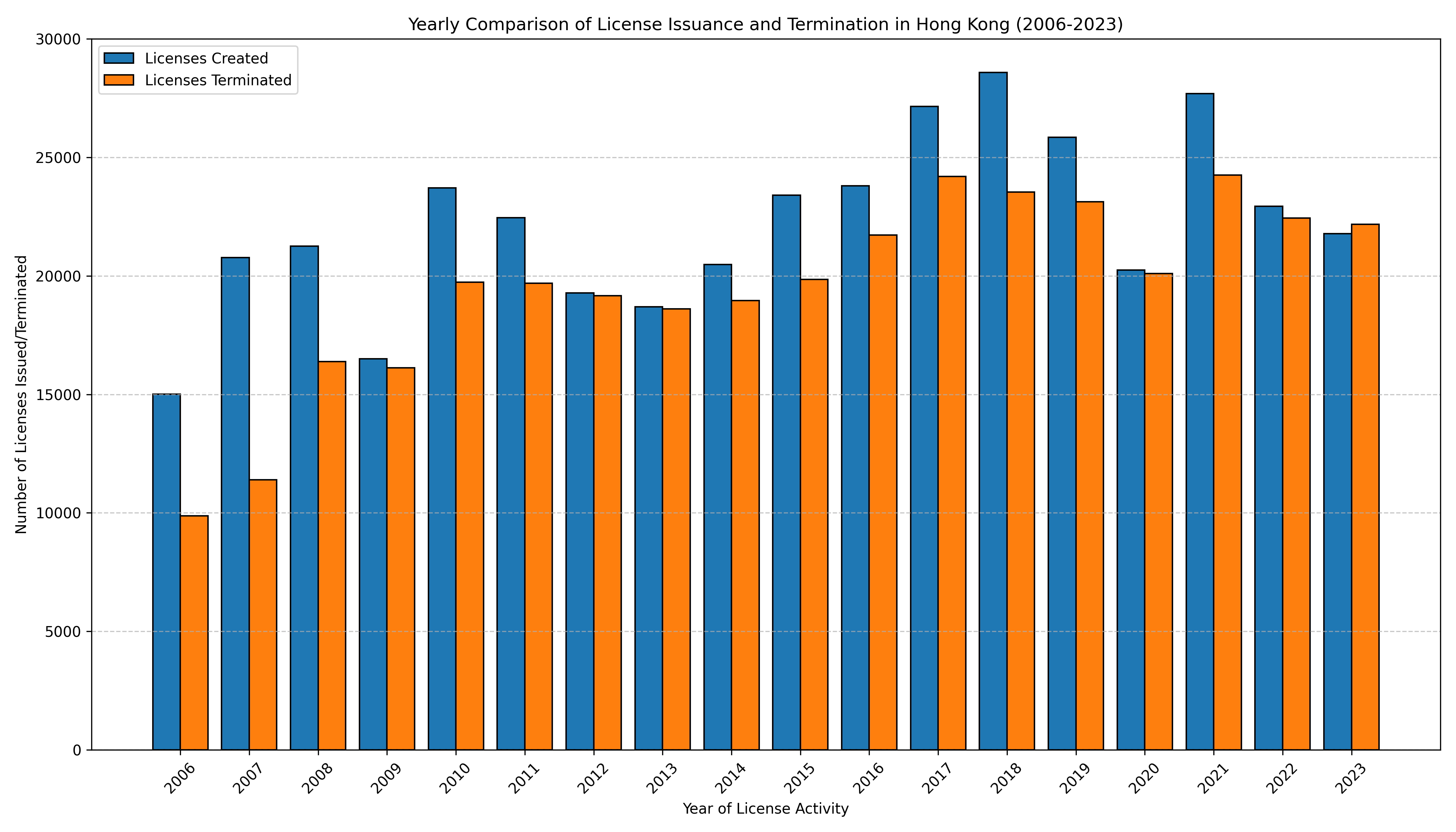}
\caption{Temporal dynamics of license creation (blue) and termination (orange), 2006–2023. 
Most years exhibit net industry expansion, with new licenses exceeding terminations. 
During the 2009 Global Financial Crisis and the 2020 COVID-19 pandemic, however, issuance and termination reached parity, indicating stagnation and stress in the financial labor market.}
\label{fig:turnover_dynamics}
\end{figure}

Table~\ref{tab:dataset_stats} reports key characteristics of the dataset. The median tenure of 1.5 years, together with the high annual turnover rate (above 25\%), highlights the volatility of employment in Hong Kong’s financial sector. Figure~\ref{fig:turnover_dynamics} shows the temporal evolution of hires and departures. 
In most years, new licenses exceed terminations, reflecting steady industry expansion. 
However, during systemic shocks such as the 2009 Global Financial Crisis and the 2020 COVID-19 pandemic, issuance and termination were nearly equal, indicating a halt in net job creation and highlighting periods of labor market stress that serve as natural stress tests for our models.

\begin{table}[h]
\centering
\caption{Key statistics of the SFC register (2003--2024).}
\label{tab:dataset_stats}
\begin{tabular}{l r}
\hline
Total professionals & 121{,}883 \\
Total firms & 4{,}979 \\
Employment records & 519{,}860 \\
Median license tenure & 1.5 years \\
Monthly turnover rate & 2.3\% \\
Median number of employees per firm & 21 \\
\hline
\end{tabular}
\end{table}

\section{Network Construction}\label{sec:network}

\subsection{Employee--Employee Network}
We construct temporal employee networks where nodes represent licensed professionals and edges represent co-employment. Two employees are connected at time $t$ if they are simultaneously active at the same firm. Edge weights are proportional to the overlap in employment duration, normalized by career length to account for seniority differences. 

This network captures peer relationships formed through shared workplaces. As expected, the resulting graphs exhibit strong small-world characteristics, with average path lengths around 2--3 and clustering coefficients far above those of random graphs of similar size. Degree distributions follow a heavy-tailed pattern, reflecting the presence of highly connected professionals who have worked across multiple institutions. Community detection using the Louvain algorithm reveals clusters aligned with major industry segments such as investment banking, brokerage, and asset management.

\subsection{Firm--Firm Network}
We also build firm-level networks where nodes are institutions and edges represent shared employees. Two firms are linked if at least one professional has worked at both. Edge weights can be defined in several ways, including absolute counts of shared employees, normalized overlap, or recency-weighted flows. 

This network encodes organizational proximity in the labor market. In February 2024, for example, the firm network comprised over 3,000 active nodes and roughly 95,000 edges, forming a single giant connected component. The network is scale-free with a handful of global hubs (e.g., Morgan Stanley Asia, Goldman Sachs, Citigroup), each connected to hundreds of smaller firms. The clustering coefficient is significantly higher than in an Erdős--Rényi baseline, indicating the presence of industry sub-communities.

\subsection{Temporal Dynamics}
Both networks are constructed for each monthly snapshot from 2007--2024, yielding a dynamic sequence of graphs. This temporal perspective is essential for modeling turnover: professional ties weaken over time if not reinforced, while firms continuously enter, exit, or reorganize. 

By integrating network evolution with feature propagation (detailed in Section~\ref{sec:methodology}), we capture how peer departures and organizational instability spread through the labor market. This provides the foundation for quantifying contagion effects in employee turnover.

\section{Methodology}\label{sec:methodology}

\subsection{Feature Propagation Framework}
To quantify peer influence and organizational stability, we apply a graph-based feature propagation procedure on the temporal networks. For each monthly snapshot, features are propagated across neighbors according to weighted averages: 

\begin{equation}
f'_i = \frac{\sum_{j \in N(i)} w_{ij} \cdot f_j}{\sum_{j \in N(i)} w_{ij}},
\end{equation}

where $f_j$ is the original feature of neighbor $j$, $w_{ij}$ the edge weight, and $N(i)$ the neighborhood of node $i$. This process transfers information about employee mobility, tenure, and firm stability through the network, capturing contagion effects beyond local attributes.

More generally, the propagation can be implemented as an iterative procedure across $k$ steps, as described in Algorithm~\ref{alg:propagation}. This formulation allows repeated diffusion of features through neighborhoods, although in practice we find that a single iteration ($k=1$) captures most of the predictive signal while avoiding oversmoothing.

\begin{algorithm}
\caption{Network Feature Propagation}
\label{alg:propagation}
\begin{algorithmic}[1]
\Require Network \(G = (V, E, W)\), Original features \(\mathbf{F}\), Iterations \(k\)
\Ensure Propagated features \(\mathbf{F}'\)
\State \(\mathbf{F}^{(0)} \leftarrow \mathbf{F}\)
\For{\(i = 1\) to \(k\)}
    \For{each node \(v \in V\)}
        \State \(\mathcal{N}(v) \leftarrow \text{neighbors of } v\)
        \For{each feature \(f\)}
            \State \(\mathbf{F}'^{(i)}_v[f] \leftarrow \frac{\sum_{u \in \mathcal{N}(v)} w_{vu} \cdot \mathbf{F}^{(i-1)}_u[f]}{\sum_{u \in \mathcal{N}(v)} w_{vu}}\)
        \EndFor
    \EndFor
\EndFor
\State \Return \(\mathbf{F}'^{(k)}\)
\end{algorithmic}
\end{algorithm}

In practice, we apply this propagation to both employee--employee and firm--firm networks, generating new features that reflect peer mobility, organizational stability, and systemic influences. These network-derived signals are then integrated with individual and firm-level attributes in the predictive models.


\subsection{Predictive Modeling Setup}
We train predictive models to estimate the probability of turnover in the subsequent month. To ensure temporal validity, we adopt a strict walk-forward validation strategy spanning 169 monthly periods (2010--2024). At each step, models are trained on all past data and evaluated on the next month, with a one-month gap to prevent leakage. This results in 5.5 million individual predictions across varying market conditions.

The severe class imbalance (2.3\% turnover events) is handled through undersampling of the majority class and probability calibration via isotonic regression. We evaluate two ensemble methods well-suited for tabular data:
\begin{itemize}
    \item \textbf{Random Forest}: used as a baseline, capturing non-linear interactions but without temporal propagation.
    \item \textbf{LightGBM}: a gradient boosting method optimized for efficiency, used as our main model with hyperparameters tuned for depth, learning rate, and subsampling.
\end{itemize}

\subsection{Link to Network Contagion}
This framework allows us to test whether turnover is predictable beyond random chance, and whether network-derived features add explanatory power over individual and firm-level attributes. In particular, it enables the quantification of contagion effects: if a significant portion of a professional’s network departs, does their own probability of leaving increase? This is the central research question addressed in the results.

\section{Results}\label{sec:results}

\subsection{Overall Predictive Performance}
Across 169 walk-forward evaluations (2010–2024; 5.5M predictions), the network-augmented \textsc{LightGBM} model achieves an Average Precision (AP) of \textbf{0.0384} and an AUC of \textbf{0.6303}, outperforming a non-network baseline by \textbf{+30.2\%} AP and \textbf{+9.1\%} AUC (Table~\ref{tab:model_performance_results}). Performance varies with market conditions but remains stable (AP coefficient of variation \(\approx\) 32\%).

\begin{table}[h]
\centering
\caption{Predictive performance (mean \(\pm\) sd) over 169 monthly test periods.}
\label{tab:model_performance_results}
\begin{tabular}{lccc}
\toprule
\textbf{Model} & \textbf{AP} & \textbf{AUC} & \textbf{F1} \\
\midrule
\textsc{LightGBM} (with network) & \underline{0.0384} & \underline{0.6303} & \underline{0.0649} \\
\textsc{Random Forest} (with network) & 0.0356 & 0.6146 & 0.0586 \\
\textsc{Random Forest} (no network) & 0.0295 & 0.5777 & 0.0389 \\
\midrule
\textbf{Improvement vs. no-network} & \textbf{+30.2\%} & \textbf{+9.1\%} & \textbf{+26.8\%} \\
\bottomrule
\end{tabular}
\end{table}

\noindent



\subsection{Feature Importance Analysis}

Feature attribution consistently shows that \textbf{firm-level stability} and \textbf{network-propagated signals} dominate predictive performance. Aggregated across all models, firm-level features account for \textbf{43.6\%} of total importance, network-propagated features \textbf{39.6\%}, individual employment history \textbf{12.5\%}, and demographics only \textbf{4.3\%} (Table~\ref{tab:category_importance_results}). 

Demographic attributes, enriched through LLM-based inference \cite{alnuaimi2024enriching}, account for less than 4.3\% of importance. This indicates that turnover dynamics are not driven by gender or nationality, but primarily by organizational stability and peer contagion effects. In other words, while demographics were considered, their negligible contribution reinforces the dominance of structural and relational drivers of employee exits.

\begin{table}[h]
\centering
\caption{Aggregated feature importance by category. Bars illustrate relative contributions.}
\label{tab:category_importance_results}
\begin{tabular}{l l r}
\toprule
\textbf{Category} & \textbf{Visual} & \textbf{Share} \\
\midrule
Firm-level characteristics & \rule{4.36cm}{2pt} & 43.6\% \\
Network-propagated features & \rule{3.96cm}{2pt} & 39.6\% \\
Individual employment history & \rule{1.25cm}{2pt} & 12.5\% \\
Demographic attributes & \rule{0.43cm}{2pt} & 4.3\% \\
\bottomrule
\end{tabular}
\end{table}

\begin{figure}[h]
\centering
\begin{subfigure}[b]{0.49\textwidth}
    \centering
    \includegraphics[width=\textwidth]{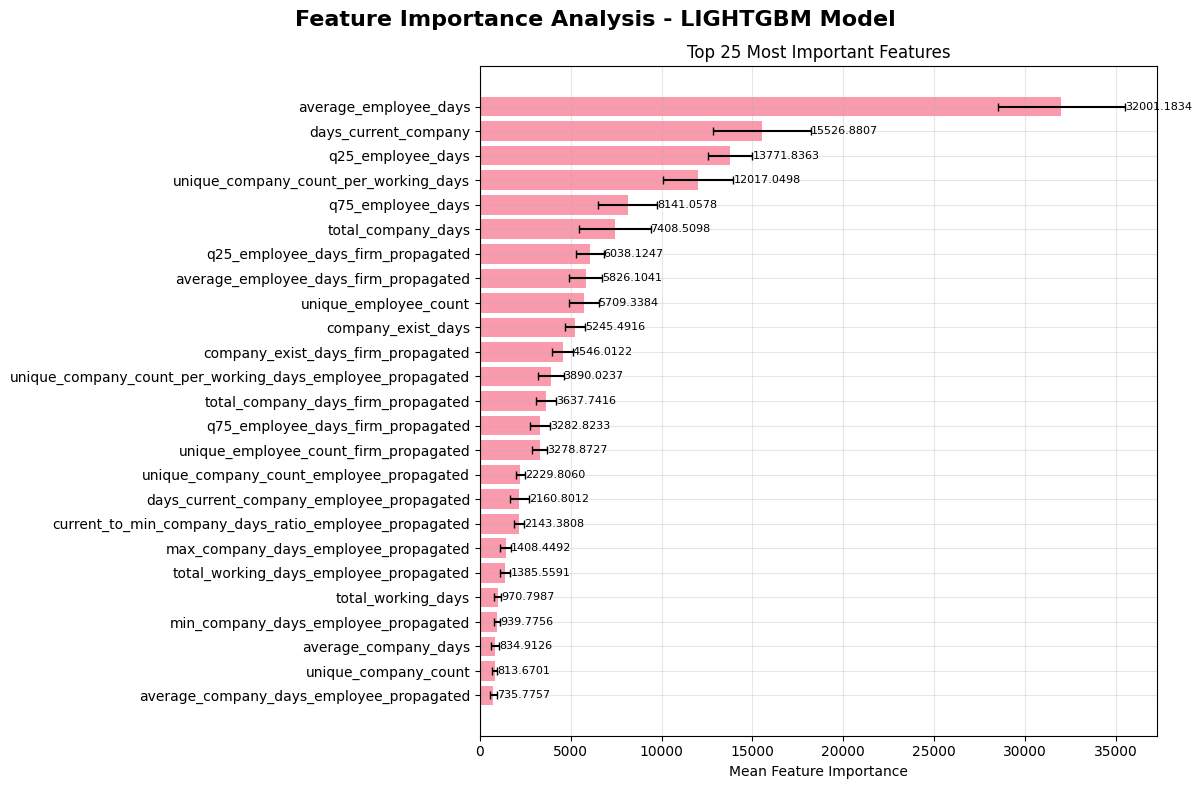}
    \caption{LightGBM (top 25 features).}
    \label{fig:lightgbm_importance}
\end{subfigure}
\hfill
\begin{subfigure}[b]{0.49\textwidth}
    \centering
    \includegraphics[width=\textwidth]{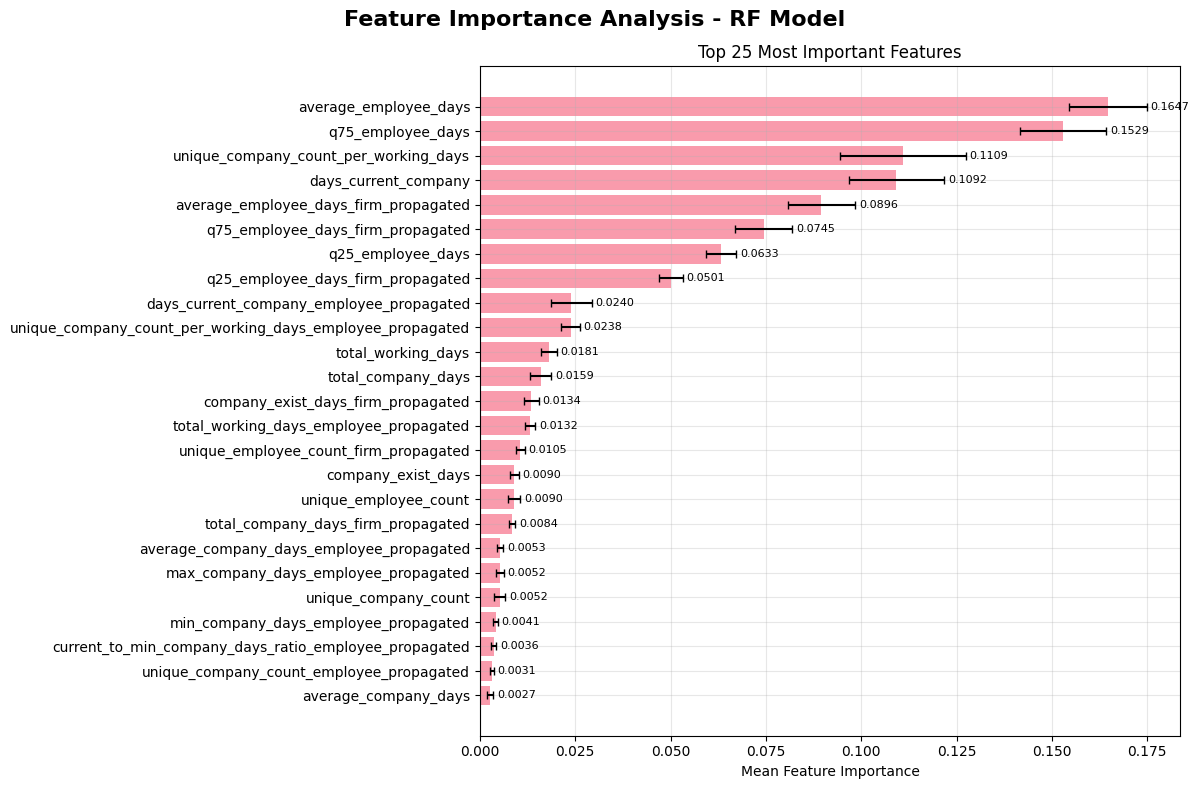}
    \caption{Random Forest (top 25 features).}
    \label{fig:rf_importance}
\end{subfigure}
\caption{Comparison of feature importance rankings across models. Both highlight the dominance of firm-level stability and network-propagated features, while demographic attributes contribute negligibly.}
\label{fig:feature_importance_models}
\end{figure}

At the feature level, the top contributors include average employee tenure, upper quantiles of employee days (stability signals), and their propagated counterparts that capture peer and industry-level dynamics. Figure~\ref{fig:feature_importance_models} compares the top 25 features as ranked by LightGBM and Random Forest models.

\subsection{Network Contagion and Threshold Effects}
We quantify peer effects by conditioning on recent departures in a professional’s immediate network (employee–employee graph, six-month window). When more than \textbf{30\%} of a professional’s peers have departed within the past six months, the individual’s turnover probability increases by \textbf{23\%} relative to baseline. 



\subsection{Temporal Heterogeneity}
\begin{figure}[h]
\centering
\includegraphics[width=0.9\linewidth]{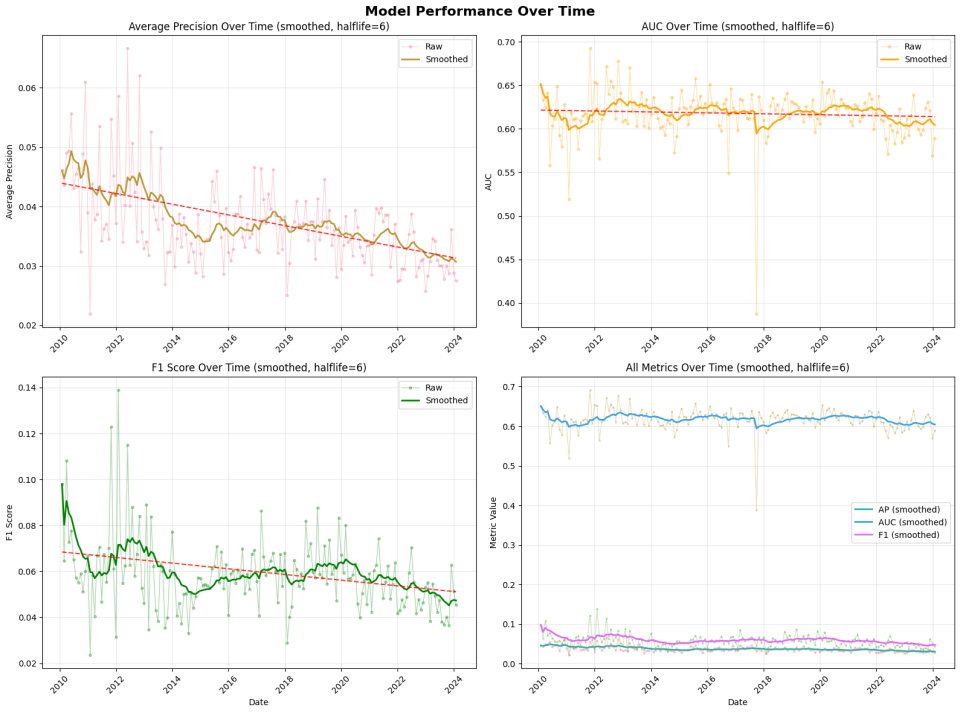}
\caption{Smoothed average precision (AP), AUC, F1 Score of the random forest model over time, 2010–2024.}
\label{fig:ap_over_time}
\end{figure}

Predictability declines over time (Figure~\ref{fig:ap_over_time}). AP peaks during post-crisis years (2010–2012: \(\sim\)0.045) and decreases toward 2023–2024 (\(\sim\)0.031). This pattern is consistent with a \emph{maturing} labor market in which shocks are less synchronized and departures less clustered, reducing the strength of exploitable network signals.

\subsection{Robustness Checks}

We verify that gains are not artifacts of imbalance handling or leakage: (i) undersampling ratios around the chosen setting yield similar AP (within \(\pm\)0.002); (ii) adding a one-month gap between train and test prevents temporal leakage (removing it inflates AP and is therefore avoided); (iii) isotonic calibration improves probability quality (Brier score decrease) without materially changing ranking metrics. 

Importantly, \textbf{employee-network propagation} provides more unique information than firm-network propagation (lower correlations with originals), aligning with the intuition that peer decisions transmit more granular signals than organization-level aggregates.

\section{Conclusion}\label{sec:conclusion}

This paper demonstrates how temporal network analysis can be applied to predict employee turnover in Hong Kong’s financial sector. By constructing monthly networks of 121,883 professionals across 4,979 firms and propagating organizational and peer features, we achieve consistent improvements in predictive performance over non-network baselines. Our findings highlight three main insights: (i) professional turnover exhibits contagion-like dynamics, with a 23\% increase in departure probability once 30\% of peers leave; (ii) firm stability and network-propagated features dominate individual attributes, underscoring the structural rather than personal nature of turnover; and (iii) predictability has declined over time, consistent with a maturing labor market less prone to synchronized shocks.

Beyond methodological contributions, these results carry practical implications for regulation, workforce management, and policy design. Regulators can use network-informed risk scores to anticipate instability, firms can identify vulnerable teams for retention efforts, and policymakers can better calibrate skilled worker immigration programs. 

Future work should explore richer network modeling approaches, including graph neural networks, multilayer representations (e.g., employment, education, social ties), and cross-market extensions to compare financial centers such as Singapore or London. By integrating network science and predictive analytics, this research contributes to the growing field of workforce analytics and illustrates how complex networks can illuminate the dynamics of modern financial labor markets.

\bibliographystyle{spmpsci} 
\bibliography{main} 

\end{document}